\documentclass[authoryear,12pt,5p,times]{elsarticle}

\usepackage{aas_macros}
\usepackage[utf8x]{inputenc}
\usepackage{graphicx}
\usepackage{natbib}
\usepackage{enumerate}
\usepackage{url}
\usepackage[breaklinks=true]{hyperref}
\usepackage{breakcites}
\usepackage{microtype}
\usepackage{bm}
\usepackage{tikz}
\usepackage{color}
\usepackage{txfonts}
\usepackage{booktabs}
\usepackage{verbatim}
\usepackage{todonotes}
\usepackage{listings}
\usepackage{longtable}

\newcommand{\photz}{photo-\emph{z}}
\newcommand{\photzs}{photo-\emph{z}s}

\begin{document}


\begin{frontmatter}

\title{The Overlooked Potential of Generalized Linear Models in Astronomy-II: Gamma regression and photometric redshifts}

\author[MPE]{J. Elliott}
\ead{jonnyelliott@mpe.mpg.de}

\author[MTA]{R. S. de Souza}
\ead{rafael.2706@gmail.com}

\author[UL]{A. Krone-Martins}
\ead{algol@sim.ul.pt}

\author[OXZ]{E. Cameron}
\ead{dr.ewan.cameron@gmail.com}

\author[MPA]{E. E. O. Ishida}
\ead{emille@mpa-garching.mpg.de}

\author[AZ,JET]{J. Hilbe}
\ead{j.m.hilbe@gmail.com}

\author[]{for the COIN collaboration}

\address[MPE]{Max-Planck-Institut f{\"u}r extraterrestrische Physik, Giessenbachstra\ss e 1, 85748, Garching, Germany}
\address[MTA]{MTA E\"otv\"os University, EIRSA ``Lendulet'' Astrophysics Research Group, Budapest 1117, Hungary}
\address[UL]{SIM, Faculdade de Ci\^encias, Universidade de Lisboa, Ed. C8, Campo Grande, 1749-016, Lisboa, Portugal}
\address[OXZ]{Department of Zoology, University of Oxford, Tinbergen Building, South Parks Road, Oxford, OX1 3PS, United Kingdom}
\address[MPA]{Max-Planck-Institut f\"ur Astrophysik, Karl-Schwarzschild-Str. 1, 85748 Garching, Germany}
\address[AZ]{Arizona State University, 873701,Tempe, AZ 85287-3701, USA}
\address[JET]{Jet Propulsion Laboratory, 4800 Oak Grove Dr., Pasadena, CA 91109, USA}

\begin{abstract}

\noindent Machine learning techniques offer a precious tool box for use within astronomy to solve problems involving so-called \emph{big data}. They provide a means to make accurate predictions about a particular system without prior knowledge of the underlying physical processes of the data. In this article, and the companion papers of this series, we present  the set of Generalized Linear Models (GLMs) as a fast alternative method for tackling general astronomical problems, including the ones related to the machine learning paradigm. To demonstrate the applicability of GLMs to inherently positive and continuous physical observables,  we explore their use in estimating the photometric redshifts of galaxies from their multi-wavelength photometry. Using the gamma family with a log link function we predict redshifts from the \textit{PHoto-z Accuracy Testing} simulated catalogue and a subset of the \textit{Sloan Digital Sky Survey} from Data Release 10. We obtain fits that result in catastrophic outlier rates as low as $\sim1\%$ for simulated and $\sim 2\%$ for real data. Moreover, we can easily obtain such levels of precision within a matter of seconds on a normal desktop computer and with training sets that contain merely thousands of galaxies. Our software is made publicly available as a user-friendly package developed in \texttt{Python}, \texttt{R} and via an interactive web  application. This software allows users to apply a set of GLMs to their own photometric catalogues and generates publication quality plots with minimum effort. By facilitating their ease of use to the astronomical community, this paper series aims to make GLMs widely known and to encourage their implementation in future large-scale projects, such as the Large Synoptic Survey Telescope.
\end{abstract}

\end{frontmatter}

\topmargin -1.3cm


\section{Introduction}
\label{sec:intro}

Generalized Linear Models (GLMs), as introduced by \citet{nel72}, offer a well established statistical framework for robust modelling and prediction making. It allows the application of regression analysis when the observed quantity originates from an \textit{exponential family} distribution rather than a Gaussian~\citep[or Normal; e.g.,][]{Hil12, hil14}. As a result, GLMs offer a readily interpretable and physically-motivated approach (via family distributions) to machine learning (ML) that can be applied to a variety of astronomical data sets. Despite being widely used across a range of scientific disciplines,  such as biology \citep{bro93,Ahrestani2013}, medicine \citep{lindsey1999}, and economics \citep{pin98,Jong2008}, and its availability  within the overwhelming majority of contemporary statistical software packages (e.g.,\ \texttt{R}, \citealt{R}; \texttt{SAS}, \citealt{SAS}; and \texttt{STATA}, \citealt{Stata}), GLMs remain almost \textit{terra incognita}
within the astronomical community \citep{deSouza2014c}. 

One particular problem which presents itself as a candidate for the implementation of GLMs is the photometric redshift (\photz{}) estimation of galaxies. Although precise redshifts can in principle be directly determined through identification of known absorption or emission lines in the optical and/or near-infrared spectrum of each target galaxy, the  observational cost of this procedure can quickly become prohibitive for large scale surveys. The only feasible alternative in such cases is to use available multi-wavelength photometry to infer approximate \photz{}s instead, but this is not always a simple task.

There exist a plethora of different spectra emitted from galaxies throughout the Universe. Their characteristic features carry signatures from the galaxy's morphology, age, metallicity, star formation history, merging history, and a host of other confounding factors in addition to its redshift. Thus, making \photz{} estimation a far from trivial task. There exist several techniques which are commonly used to estimate redshifts from photometry and can be  divided in to: (i) template fitting techniques~\citep[e.g.,][]{benitez2000,bolzonella2000,ilbert2006}, and (ii) ML (or empirical) techniques \citep[e.g.][]{connolly1995, collister2004, wadadekar2005, miles2007, omill2011, reis2012, Krone-Martins14a}.
 In template fitting techniques, a set of synthetic spectra are determined from synthesised stellar population models for a given set of metallicities, star formation histories and initial mass functions, among other properties. The \photz{} is calculated by determining the synthetic photometry (and thus spectral template and redshift) which best fits the photometric observations. ML techniques, on the other hand, usually require a data set with spectroscopically measured redshifts to train the chosen method.

Many studies have examined the individual advantages of each \photz{} code~\citep[for a glimpse on the diversity of existent methods, see][and references therein]{hildebrandt2010, abdalla2008, zheng2012, Sanchez14a}. \citealt{abdalla2008} investigated the differences between five commonly used template fitting codes and a neural network. The neural network proved to be more reliable in redshift ranges with a higher density of training data, while the template fitting methods depended heavily on the underlying templates. Despite these caveats, the overall performance of all codes were, to first order, consistent and displayed catastrophic errors ranging from $5-9\%$, which is considered good in terms of \photz{} estimates~\citep{abdalla2008}. More recently, methods which combine several \photz{} techniques in a Bayesian approach, coined ensemble learning, have begun to be implemented with the hope that they can complement each other{\bf '}s drawbacks~\citep{Carrasco14a}.

One of the largest practical difficulties for the current \photz{} methods is the time necessary to either fit the templates or train the underlying ML method; on top of that, the required size of the training set is often highly influential for empirical methods~\citep[][]{Firth03a}. \emph{Big data} catalogues expected from large sky surveys, like the \textit{Large Synoptic Survey Telescope}\footnote{\url{http://www.lsst.org/lsst}} \citep{lsst},  \textit{EUCLID}\footnote{\url{http://sci.esa.int/euclid}} \citep{euclid} or the \textit{Wide-Field Survey Infrared Telescope}\footnote{\url{http://wfirst.gsfc.nasa.gov}} \citep{wfirst}, warrant the need for fast and reliable \photz{} methods that are capable of processing such large volumes of data in minutes to days rather than years,  thereby facilitating higher level analyses and model refinements for downstream data products.

In this work, we introduce a new technique based on robust principal component analysis (PCA) and GLMs to estimate \photz{}s. The method runs in a matter of seconds on a single core computer, even for millions of objects. In addition, we achieved very low levels of  catastrophic errors when using training sets of a few thousands objects. The combination of short computational run time, moderate training set size, and small catastrophic errors makes GLMs a robust and implementable technique for future large scale surveys. 

The outline of this article is as follows. In \S \ref{sec:method}, we give a broad overview of GLMs, in \S \ref{sec:data} we provide a description of the data set utilised. The methodology implemented is outlined in \S \ref{sec:imp}. We then present our results and compare with the recent literature in \S \ref{sec:res} and summarise our conclusions in \S \ref{sec:conclusion}.



\section{Overview of Regression Methods}
\label{sec:method}

Before we delve into the details of GLMs and the gamma family, we make a brief overview of linear regression, a common tool used  within astrophysics. Afterwards, we explicitly outline the details of GLMs with the gamma family and explain how it can be applied to determine \photz{}s for a particular  data set. 

\subsection{Overview of Linear Regression}

Consider a given data set containing $N$ (distinct objects; e.g., galaxies), \[\mathcal{D}=\{(x_1, y_1), (x_2,y_2),...,(x_N,y_N)\},\] where the $x_i$ are observations of the independent variable, $X$, and the $y_i$  are observed values of a dependent random variable (RV), $Y$, which is a function of $X$, $Y=f(X)$. Traditionally, $X$ is called the \textit{explanatory variable} and $Y$  the \textit{response variable}. The expected value and variance of $Y$ are denoted by E($Y$) and var($Y$), respectively. In this context, a linear model describes the response variable \textbf{($Y$)} as a linear function of the explanatory variable \textbf{($X$)}:
\begin{equation}
Y = \beta_0 + \beta_1 X + \epsilon = \eta + \epsilon,
\label{eq:linmod}
\end{equation}
where $\{\beta_0,\beta_1\}$ are scalars called \textit{slope coefficients} or \textit{covariates}, $\eta=\beta_0+\beta_1 X$ is the linear component (or \textit{predictor}) of this simple model. Finally, $\epsilon$ is an error term considered to be independent and identically distributed, $\epsilon \sim N(0,\sigma^2)$.

When a standard linear regression approach is applied, the linear predictor in equation \ref{eq:linmod} is assumed to fully describe the response variable. The measured values are used to determine the covariates of the linear predictor that uniquely identify a straight line through the chosen data set minimising the error term. Having the scalar coefficients determined, the model provides a direct relation between $X$ and $Y$, allowing one to predict the mean value of $Y$ for a given measurement of $X$.

In order to clarify the procedure described in the next subsections, we invite the reader to approach this simple linear regression problem from an alternative  perspective. Consider now each measurement, $\{x_i, y_i\}$, as a realisation of different variables $\{X_i, Y_i\}$ from a common family of probability density functions (PDFs), but with distinct parameters $\mu_i$ for each index $i$. The underlying PDF driving the behaviour of the response variable ($Y_i$) will be denoted by $f(y_i;\kappa_i)$,  where $\kappa_i$ is the parameter vector of the PDF underlying the $i$-th measurement. If $Y_i$ follows a Normal PDF with mean $\mu_i$ and variance $\sigma_i^2$, then
\begin{equation}
f(y_i;\kappa_i)=\frac{1}{\sqrt{2\pi\sigma_i^2}}\exp\left[-\frac{1}{2}\frac{\left(y_i-\mu_i\right)^2}{\sigma_i^2}\right],
\label{eq:PDF}
\end{equation}
where $\kappa_i=\{\mu_i, \sigma_i\}$. This is summarised as $Y_i\sim N(\mu_i, \sigma_i)$. For reasons which will be clarified later, we consider $\sigma_i$ a fixed value and, thus, determining $\mu_i$ is enough to completely characterise $f(y_i;\kappa_i)$. In this context, we can relate the measured $x_i$ to the expected value of the corresponding response variable, $y_i$=E($Y_i$), though the slope coefficients\footnote{If our model is correct, for each measurement the expected value should correspond on average to the measured one.}
:
\begin{equation}
y_i=\textrm{E}(Y_i)=\mu_i=\textbf{x}_i^T\bm{\beta},
\end{equation}
with $\textbf{x}_i=\{1, x_i\}$ and $\bm{\beta}=\{\beta_0, \beta_1\}$. We can now use the chosen PDF family (e.g., equation \ref{eq:PDF}) and the observed data $\mathcal{D}$ in order to find values of $\bm{\beta}$  which better describes the data. In this case, the model is composed of two main ingredients:
\begin{enumerate}[1.]
\item a PDF underlying the behaviour of each response variable, $f(y_i;\mu_i)$ and
\item a relation connecting E$(Y_i)$ with measured values of the explanatory variable, also called the \textit{link function}, $g(\textrm{E}(Y_i))=\mu_i$. 
\end{enumerate}
Once the values of parameter $\bm{\beta}$ are determined, we can use the inverse link, $g^{-1}$, to calculate the expected value for the response variable given a measured  input $x$. 

Note, that in this simple example, although the mathematical expressions in both approaches are the same, their interpretations are different. In the standard analysis, the best-fitted parameters define completely the connection between $X$ and $Y$. In the second approach, they characterize a linear relationship between measurements of $X_i$ and a parameter  which uniquely identifies a PDF underlying each response variable $Y_i$, $\kappa_i$. This approach allows us to extend the same reasoning to situations where a linear relation is not a good description of the process driving the data.
 
This example can be generalised for the case with more than one explanatory variable, forming a multiple linear model. The \textit{general linear model}\footnote{Not to be confused with General\textit{ized} Linear Models (GLMs).}  goes  one step further, including situations where there are more than one response variable. In such models the errors are considered to be uncorrelated and follow a multivariate normal distribution. Although a very useful tool, general linear models are not suited for situations where there are restrictions on $Y$ (e.g., binary, count or strictly positive data) or when the variance depends on the mean. The GLMs are a generalisation of this framework, capable of handling both scenarios.

\subsection{Generalized Linear Models}
\label{subsec:GLMs}

This subsection presents an overview of a vast subject. For a detailed theoretical review see \citet{dobson2002} and \citet{Hil12}.
 
The example from the previous section clearly shows that the PDF is a key ingredient in the construction of a GLM framework. Moreover, the procedure relies on one main feature of the PDF:  within the chosen family, a distribution should be uniquely identified through one single parameter $\mu$ (called \textit{location} or \textit{mean}). Determining this parameter is the ultimate goal of the GLM methodology. 

In order to fulfil this requirement, GLMs are constructed for any distribution that belongs to the \textit{exponential family} of distributions (Gaussian/normal, gamma, inverse Gaussian, Bernoulli, binomial, Poisson, and negative binomial). The PDF for any member of this family can be written as

\begin{equation}
f(y;\theta) = s(y)t(\theta)e^{a(y)b(\theta)},
\end{equation}
where $s$, $t$, $a$ and $b$ are known functions and $\theta$ is the \textit{canonical parameter}. The reader may also find them as,
\begin{equation}
  f(y;\theta)  = \exp\left[a(y)b(\theta)+c(\theta)+d(y)\right],
 \label{eq:canPDForig}
 \end{equation}
with $c(\theta)=\ln[t(\theta)]$ and $d(y)=\ln[s(y)]$. 
  When $a(y)=y$, the distribution is said to be in the \textit{canonical form}.  \textit{Natural exponential distributions} are exponential ones in the canonical form with  $b(\theta)=\theta$. If there is an extra parameter $\phi$ it is considered known\footnote{In the Gaussian example described previously, $\phi=\sigma^{2}=\textrm{var}(Y)$.}. Natural exponential distributions can always take the form

 \begin{equation}
   f(y; \theta, \phi) = \exp\left\{ \frac{y\theta-A(\theta)}{B(\phi)}+C(y;\phi) \right\} 
   \label{eq:canPDF}
 \end{equation} 
 \noindent \citep[see ][section 2.3]{Hil12}. The above expression  contains $A(\theta)$, which is called the \textit{cumulant}, $B(\phi)$ the \textit{scale parameter}, $\phi$ the \textit{dispersion parameter}, and $C(\cdot; \cdot)$, the normalisation term that scales the integral to unity. For these distributions, 
  \begin{eqnarray}
    \textrm{E}(Y)   & = & A'(\theta),  \nonumber \\
    \textrm{var}(Y) & = & A''(\theta)B(\phi).
    \label{eq:canVAR}
  \end{eqnarray}
  
\noindent In the specific normal distribution example described previously, it is simple to show that $\textrm{E}(y)=\mu$ and $\textrm{var}(Y)=\sigma^{2}$. 

Two other important ingredients in the structure of a GLM model are the linear component $\eta$ and the link function $g(\cdot)$. In the simple linear regression case, $\eta$ was equivalent to our assumption of a linear relation between the explanatory and response variables and $g$ was merely the identity function ($g(\mu)=\mu$). However, when  $Y_i$ relates to $X_i$  through a non-linear expression, $\eta$ will play the important role of linearising the connection between $X_i$ and E($Y_i$).  In other words, even if we are dealing with non-linear data, we can still define the link function as
\begin{equation}
g(\mu)=\eta=\textbf{X}^T\bm{\beta},
\label{eq:linketa}
\end{equation}
given that the PDF belongs to the one parameter exponential family.
After the determination of $\bm{\beta}$, the inverse link function is used to determine $\mu$ and from this we know that $g$ must be invertible. For PDFs in the canonical form the \textit{canonical link function} is given by $b(\theta)$ (equation \ref{eq:canPDForig}).

In summary, all GLMs share a similar structure and are characterized by:

\begin{itemize}

 \item A random response component whose mean $\mu$ is to be estimated.  The response variable, $Y$, is assumed to be theoretically derived as a random sample of an underlying single parameter PDF belonging to the GLM family of distributions. The goal of modelling $Y$ is to find an unbiased estimate of the mean parameter which better describes the data. 

 \item A systematic (or linear) component, $\eta$, built from the explanatory variables, $\mathbf{X}$ (sometimes called \textit{covariates}), and their associated slope coefficients, $\bm{\beta}$. Their multiplication produces a linear predictor for each observation (equation \ref{eq:linketa}).

 \item A link function, $g(\cdot)$, which defines how the mean is associated with the explanatory variable. The link function linearises the relationship between the mean response and predictors ($X_i$). Once the slope coefficients are determined, we are able to use the inverse link and the observed $X_i$ to estimate the mean, i.e.,  

\begin{equation}
  \mu = g^{-1}(\eta).
\end{equation}

\item Conversion of the PDF to a log-likelihood function for the observed data, which is used as the basis for the determination of the slope coefficients.

\end{itemize}

\subsection{Gamma Family and Regression}

The gamma distribution is characterised by the response variable, $Y$, taking only positive real values. It is optimal when fitting positive-only values with a shape determined by its estimated parameter. As a single parameter model the GLM gamma model is limited to a specific set of values. If a second scale parameter is employed, the range of shapes allowed by the model are greatly enlarged. This is not a GLM model though. On the other hand, if the GLM gamma model does fit a given data situation, the model is more efficient and easier to interpret.

In this study we predict the \photz{} of a galaxy from multi-wavelength photometry and compare it to the measured spectroscopic redshift, $z_{\rm spec}$. As redshift is always positive and continuous, we can apply a gamma family distribution, which in its exponential family form can be expressed as \citep{Hardin01a}:
\begin{eqnarray}
  f(y;\mu, \phi) & = & \frac{1}{y\Gamma(1/\phi)}\left(\frac{y}{\mu\phi}\right)^{1/\phi}\exp\left(-\frac{y}{\mu\phi}\right)\nonumber\\
                 & = & \exp\left\{\frac{y/\mu - (-\ln\mu)}{-\phi}+C(y;\phi)\right\}, \nonumber \\
\label{eqn:gamma_model}
\end{eqnarray}
\noindent where
\begin{equation}                 
  C(y, \phi)=\frac{1-\phi}{\phi}\ln y-\frac{\ln\phi}{\phi}-\ln\Gamma\left(\frac{1}{\phi}\right).
  \label{eqn:gamma_model2}
\end{equation}

From equations \ref{eq:canPDF} and \ref{eqn:gamma_model}, we recognise the canonical parameter as  $\theta=1/\mu$, the cumulant $A(\theta)=-\ln\mu$ and the scale parameter $B(\phi)=-\phi$. Consequently, from equation~\ref{eq:canVAR}, $\mathrm{E}(y)=\mu$, and $\mathrm{var}(y)=\sigma^2=\mu^{2}\phi$. It is important to note that the GLM gamma model describes a response variable as a distribution with constant coefficient of variation, $\sigma/\mu$ (standard deviation / mean). Thus, determining $\mu$ is enough to univocally identify a distribution.

A useful feature of GLMs is that a link function can be assigned to the estimation algorithm, and it does not only have to be a natural or canonical link. For example, the log link, $g(\mu)=\ln(\eta)$, is commonly associated with the gamma model. When this is employed in place of the natural link function, the associated inverse link, $\mu=\exp(\eta)$, ensures a positive mean for any $\eta$. Unless the mean and data are inversely related statisticians typically use the log link for this continuous response model.

This is also the approach chosen for this work. In what follows, we use the log-link function $\mu^T=\log\left(\bm{\beta}^T\mathbf{X}\right)$, where $\mathbf{X}$ is a $m\times n$ matrix containing  $m$ magnitudes for each of the $n$ galaxies and $\bm{\beta}$ is the covariate  column vector. The covariates are estimated by maximising the log-likelihood of the regression model (Eqn.~\ref{eqn:gamma_model}) utilising iteratively re-weighted least-squares~\citep[see][and references therein]{Hardin01a}. 

In addition, to avoid numerical instabilities and identifiability issues it is preferable for the predictor variables, $\mathbf{X}$, not to exhibit strong correlations (i.e., \textit{multicollinearity}).
This is not necessarily the case for the magnitudes of galaxies that can be strongly correlated across different broadband filters. As a precaution against multicollinearity we carry out \textit{Principal Component Analysis} and adopt the principal components (PCs) of the observed magnitude set as our explanatory variables\footnote{For a general review on PCA, see \citet{jollife2002}. For examples of PCA use in astronomy, see also e.g. \citet{conselice2006,ishida2011,ishida2011b,ishida2013,jeeson2011, Krone-Martins14b}.}. Beyond ensuring non-correlated features, using the PCs also optimises the use of computational resource{\bf s}, as the calculation time required increases non-linearly with the number of explanatory variables. As a dimensionality reduction technique, PCA allows a robust way to increase the execution speed of the redshift estimation.

As a final remark, we emphasise that the key to understanding a statistical model is to determine how well the predicted values fit the observed ones. Each type of model, being based on a specific probability distribution, or mixture of distributions, is limited to a specific range of predicted values. That is, the distributional assumptions of a gamma distribution determine the range and shape of predicted values it can have. If the observed response values differ greatly from the possible predicted values, the model cannot fit the data well. In such a case it is important to find another model more suited to fitting the observed values. In the next sections we quantify the ability of the gamma family GLM in predicting \photz{}s of galaxies by comparison with simulations and measured spectroscopic redshifts.



\section{Data}
\label{sec:data}

To compare GLMs with other methods and to better put our results in to context, we adopted a publicly available data set that was previously submitted to different \photz{} codes. The \textit{PHoto-z Accuracy Testing} (PHAT) was an international initiative to identify the most promising \photz{} methods and guide future improvements. Two observational photometric catalogues were provided: PHAT0 with simulations, and PHAT1 with real observations. A total of 17 \photz{} codes where submitted. As a direct comparison using PHAT1 is not possible, since the answers of the challenge are not openly available, we applied GLMs to PHAT0 and compare its results to those reported by \citet{hildebrandt2010}. PHAT0 has $169,520$ simulated galaxies with redshifts ranging from $z=0.02-2.24$, and magnitudes in 11 filters ($u$, $g$, $r$, $i$, $z$, $Y$, $J$, $H$, $K$, $IRAC1$, and $IRAC2$).

In addition, we apply the same technique to a real data set obtained from the {\it Sloan Digital Sky Survey} \citep[SDSS;][]{York00a}. We obtain a galaxy sample using the same selection criteria and SQL query (\ref{app:sql}) outlined in \citealt{Krone-Martins14a} from the most recent Data Release 10 \citep{Ahn14a}. This results in a sample of $1,347,640$ galaxies with a redshift range of $z=0-1.0$, with magnitudes in 5 filters ($u'$, $g'$, $r'$, $i'$, and $z'$). To compare the same data set, but with dereddened magnitudes, we also use the SQL query outlined in \citealt{Carrasco14a}.



\section{Methodology}
\label{sec:imp}

On the basis of making GLM modelling easily accessible to the community, we have developed a set of \photz{} packages which can be used on any multi-wavelength data set~\citep{desouza14b}. The codes are written in both \texttt{R}\footnote{\url{www.r-project.org/}} (\ref{app:R}), a programming language commonly used in the statistical sciences, and \texttt{Python}\footnote{\url{www.python.org/}} (\ref{app:python}) which  is becomingly widely adopted in the astronomical community. In addition, we have implemented a web application using the \texttt{Shiny}\footnote{\url{shiny.rstudio.com/}} (\ref{app:shiny}) platform whereby users can upload their data set and have the \photzs{} and diagnostic plots delivered. We adopted the following step-by-step methodology to determine the \photzs{} of a sample of galaxies using their multi-wavelength photometry.

\begin{figure}
   \includegraphics[width=9cm]{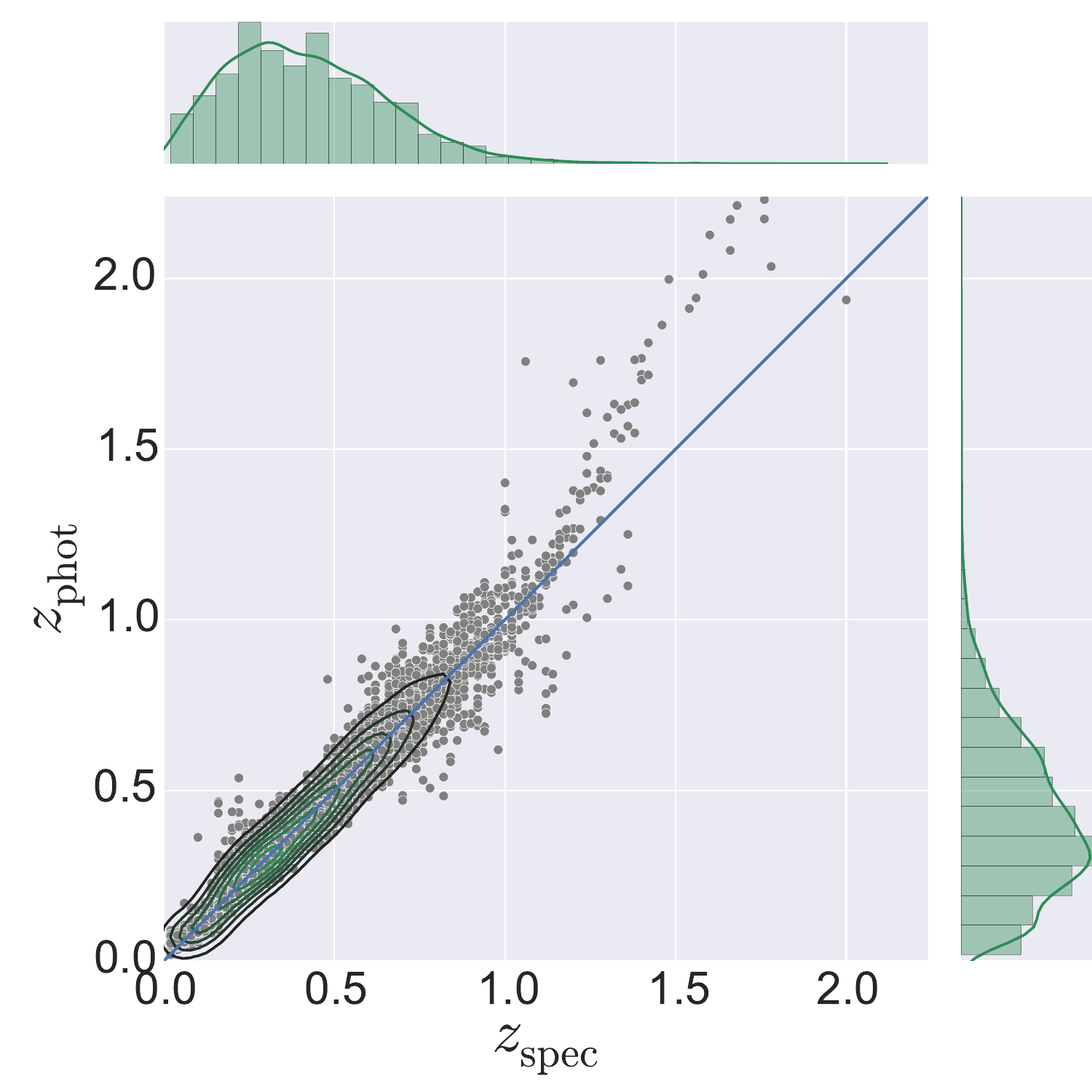}
   \includegraphics[width=9cm]{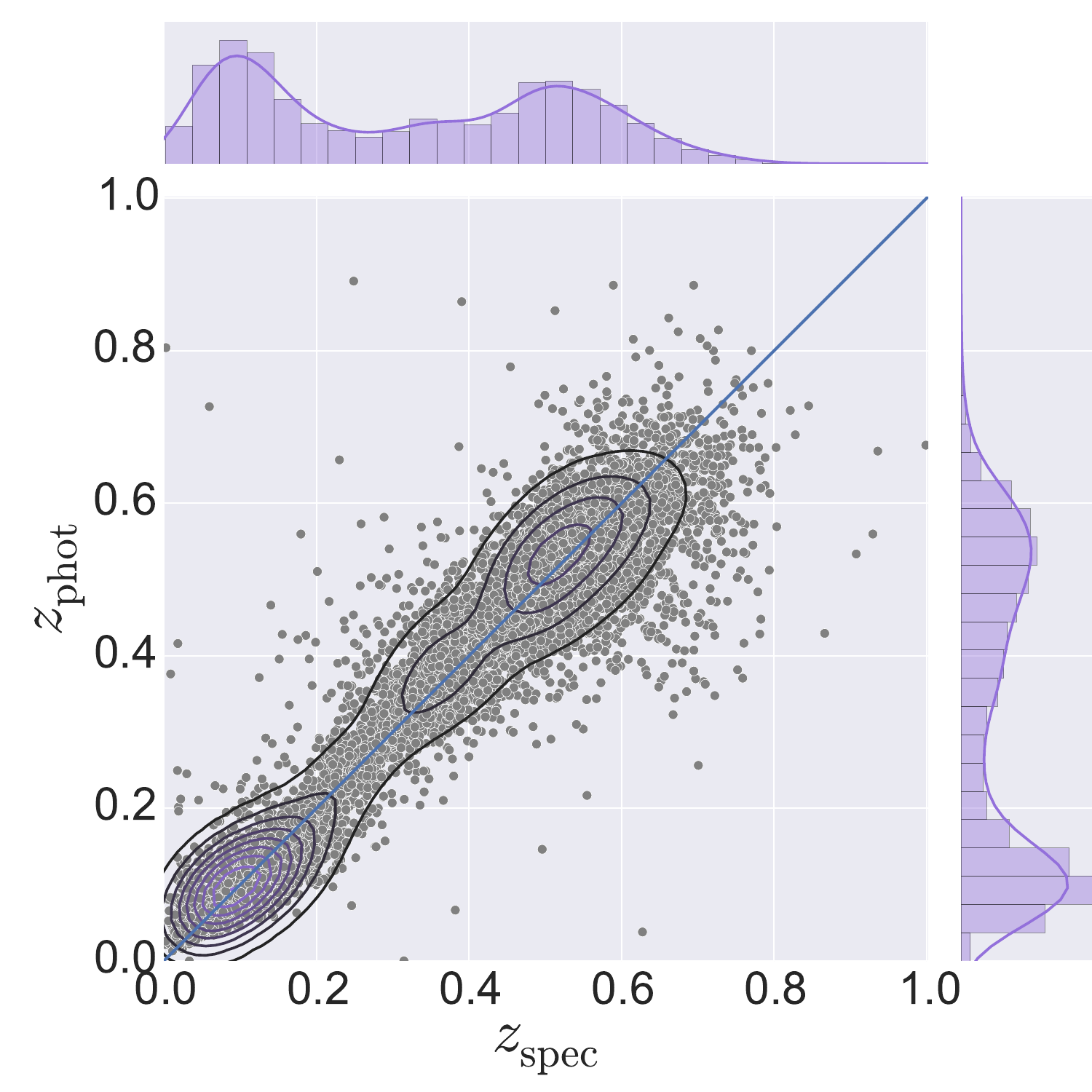}
   \caption{The 2D probability density of the predicted redshift from the GLM fit vs. the spectroscopic redshift (central plots). The upper and right subplots in each panel depict the redshift distribution along \photz{} and $z_{\rm spec}$, respectively. We note that this is a randomised subsample amounting to 10,000 galaxies. {\bf Top}: Results for the PHAT0 sample (green). {\bf Bottom}: Results from the SDSS sample (purple).
   }
  \label{fig:kde_plot_2d}
\end{figure}

\begin{enumerate}[1.]

  \item The data was randomly split into training and test sets with the training sample holding at least $10\%$ of the number of galaxies (see section \ref{subsec:caveats} for a detailed analysis on the influence of this choice in our final results.). 
  
  \item Robust principal component analysis (e.g., \citealt{Candes11a,desouza2014}) was carried out on the complete data set, training and test, to ensure the PCs are not dominated by one of the two samples. We note that this corresponds to a semi-supervised technique since, that data without any measured redshifts can help determine the PCs \citep[see, e.g.,][]{shah2008svm}. The threshold on cumulative percentage of total variance was set to $\sim 99.5\%$ in order to determine the number of PCs to be used with the GLM. 
 
  \item We utilised a gamma family distribution to reflect the fact that measured redshifts are positive and continuous. The relationship between the redshift and the explanatory variables (our linear predictor; equation~\ref{eq:linketa}) took the \texttt{R} formula form of 
  \begin{equation}
    z_{\rm phot} \sim PC_{1}^2 + PC_{1}*PC_{2}*....*PC_{n} + \rm C, 
    \label{eqn:formula}
  \end{equation}
  \noindent where $n$ is the number of principal components and $C$ is a constant. \texttt{R} formula of this kind are called \textit{simple factorial model formulae}, where the $*$ is a \textit{crossing} operator, which allows the inclusion of interaction terms. For example, $A*B=A+B+A\cdot B$~\citep[for a full description see,][]{nel65}. This formula was fit using iteratively re-weighted least squares. We note that tests using a simple formulation without interaction terms are inadequate in capturing the complexity of the data.
  
  \item The predicted \photz{} for the test data was calculated using the principal component projections of the test data set and the best-fit GLM using the training sample.
  
  \item To measure how well the \photzs{} were estimated, we employed a metric commonly used in the literature, specifically, the catastrophic error (or outlier rate/fraction). We applied two definitions of this metric: one used for the PHAT0 sample by \citealt{hildebrandt2010}, and the second used more commonly in the literature and the PHAT1 sample by \citealt{hildebrandt2010}. We define them respectively as:
  {
  \begin{enumerate}
          \item {\begin{equation}
                 O^{(a)} = \Delta z^{(a)} > 0.10,
                 \end{equation}
         \noindent where $\Delta z^{(a)} = |z_{\rm phot}-z_{\rm spec}|$, and
         }
         
         \item {\begin{equation}
                 O^{(b)} = \Delta z^{(b)} > 0.10,
                 \end{equation}
         \noindent where $\Delta z^{(b)} = \frac{|z_{\rm phot}-z_{\rm spec}|}{1+z_{\rm spec}}$.
         }
   \end{enumerate}
  }
  
\end{enumerate}


\begin{figure*}
   \includegraphics[width=9cm]{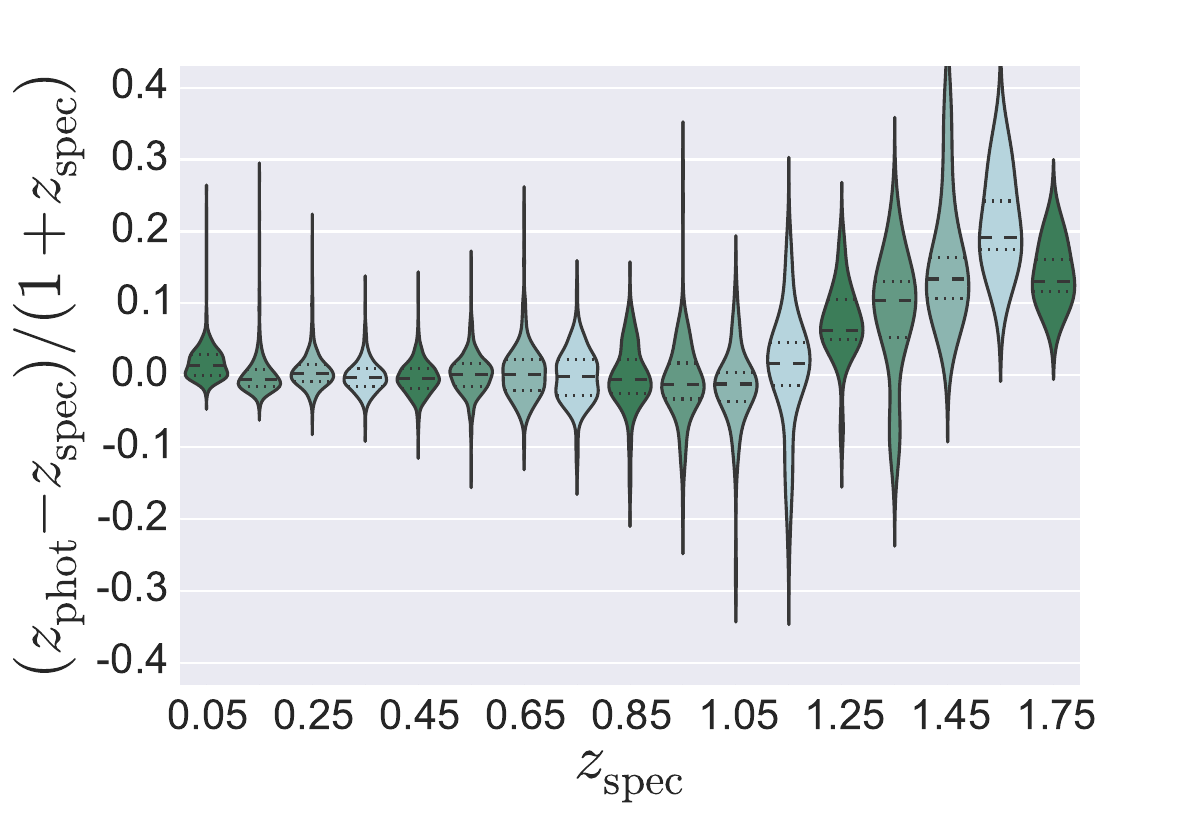}
   \includegraphics[width=9cm]{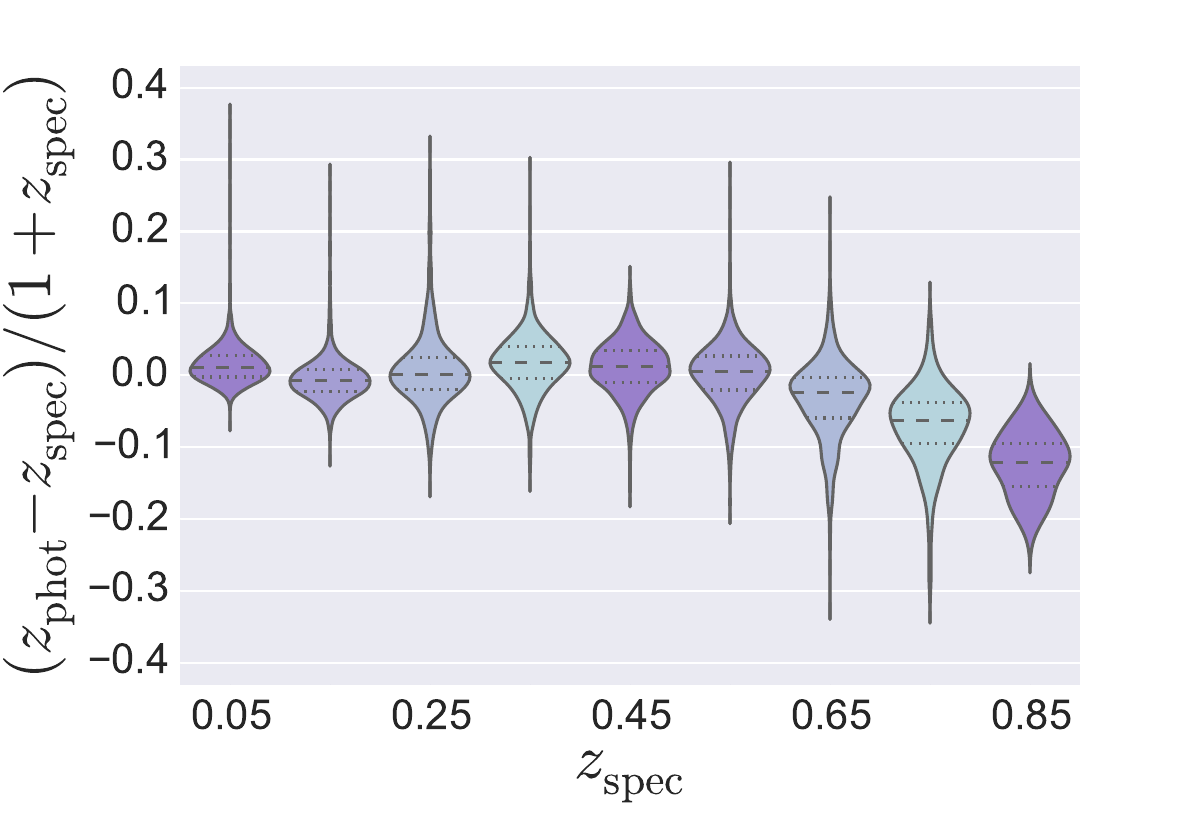}
   \caption{Violin plots that depict the probability density of \photz{} errors per redshift bin, $\Delta z_{\rm spec}=0.1$, determined from the entire galaxy sample. In each violin the central dashed line is the median and the dotted lines are the 25\% and 75\% quartiles. {\bf Left}: Results from PHAT0 sample (green). {\bf Right}: Outcomes from SDSS sample (purple).}
  \label{fig:violin_plot}
\end{figure*}



\section{Results}
\label{sec:res}

\subsection{Simulation}
\label{subsec:simulation}

The PHAT0 data set was fit using the gamma family with a log link. We split the data into training and testing with a ratio of 1:9, respectively. This criterion resulted in a training size of 33,904 galaxies and a testing set of 135,616 galaxies. Decomposition by PCA showed that at least 6 components needed to be used to ensure that $99.5\%$ of the variance was retained. The best-fit GLM is shown in Fig. \ref{fig:kde_plot_2d} (top panel) which presents a catastrophic error rate of $O^{(a)}=4.4\%$. The time taken to fit amounted to $170$ seconds on an AMD Athlon X2 Dual-Core QL-64 processor with $1.7\, \rm GB$ RAM on the Ubuntu  10.04 operating system, which represents an old laptop at today's standards. We note that changing either the training set size or the number of PCs does not dramatically alter the solution. For example, using 8 PCs ($99.95\%$ variance) results in $O^{(a)}=2.5\%$ catastrophic errors determined in $1200$ seconds or 10 PCs results in $O^{(a)}=1.4\%$ in $4948$ seconds. 

The central plot in Fig.~\ref{fig:kde_plot_2d} (top panel) shows $z_{\rm spec}$ compared with the \photz{} calculated from our GLM gamma model, while the top and right plots represent their individual distributions for the entire redshift range. Numerical diagnostics for such results are displayed in Table \ref{tab:diag}. Fig.~\ref{fig:violin_plot} (left panel) details the redshift distribution of \photz{} calculated for PHAT0 in bins of $\Delta z_{\rm spec}=0.1$. There is a tendency for a larger bias to be present in higher redshifts ranges, reflecting the characteristics of the training set. In these regions, not only does the data quality decrease significantly, but there is also an observational effect that favours brighter galaxies. Thus, the presence of such biases are not completely unexpected. However, investigating if there is a preference for under/over estimation of \photz{} within the GLM framework would require a $k$-fold cross-validation analysis which is out of the scope of this work. We do highlight though, that such a study is crucial for a potential user that aims to optimise their \photz{} results.

We note that our results give a slightly larger catastrophic error than compared with the literature. However, we highlight that the simulated data set was created from template techniques and, as such, a lower catastrophic error would suggest that the technique has results that reflect those to template fitting rather than the absolute performance of the technique~\citep{hildebrandt2010}. Finally, after demonstrating that the GLM gamma model has competitive capabilities in comparison with other techniques, we investigated its absolute ability with observed data sets.

\begin{table*}
\vspace{-2cm}
\caption{Diagnostic comparison of samples}
\begin{tabular}{p{9cm} l l l l}
\hline
Code & Type$^{a}$ & $\mathrm{bias}_{z}^{b}$ & $rms(\Delta z)^{c}$ & Outlier rate \\
\hline
& & & & $\%$ \\
\hline

PHAT0 & & & & \\
{\bf This work} & Empirical & 0.033 & 0.025  & 1.367 (10 PCs) \\
                & &      &        & 2.511 (8 PCs) \\ 
                & &      &        & 4.438 (6 PCs) \\
Le PHARE$^{d}$ & Template & 0.000 & 0.010 & 0.044 \\
Bayesian Photo-{\it z}'s$^{d}$ & Template & -0.005 & 0.011 & 0.026 \\
Easy and Accurate $Z_{\rm phot}$ from Yale$^{d}$ & Template & -0.001 & 0.012 & 0.000 \\
GALaxy EVolution and GAZELLE$^{d}$ & Template & 0.000 & 0.014 & 0.053 \\
GOODZ$^{d}$ & Template & 0.000 & 0.012 & 0.018 \\
HyperZ$^{d}$ & Template & -0.002 & 0.013 & 0.185 \\
Low-Resolution Spectral Templates$^{d}$ & Template & 0.000 & 0.011 & 0.026 \\
Purger (template repair)$^{d}$ & Template & -0.005 & 0.011 & 0.053 \\
Zurich Extragalactic Bayesian Redshift Analyzer $^{d}$ & Template & 0.000 & 0.011 & 0.062 \\
Zurich Extragalactic Bayesian Redshift Analyzer (modified)$^{d}$ & Template & -0.005 & 0.011 & 0.044 \\
Artificial Neural Network \photz{}$^{d}$ & Empirical & 0.000 & 0.011 & 0.018 \\
Boosted Decision Trees$^{d}$ & Empirical & -0.004 & 0.019 & 0.389 \\
Purger (nearest-neighbour Fit)$^{d}$ & Empirical & 0.000 & 0.017 & 0.053 \\
Polynomial fitting$^{d}$ & Empirical & 0.001 & 0.019 & 1.669 \\
Regression Trees$^{d}$ & Empirical & 0.000 & 0.013 & 0.010 \\
Singal Neural Networks$^{d}$ & Empirical & -0.005 & 0.049 & 18.202 \\

\hline\hline

SDSS & & & & \\

{\bf This work} & Empirical & 0.038 & 0.029 & 6.44 (5 PCs) \\
                &  &     &       & 8.09  (4 PCs) \\
Eureqa (Symbolic Regression)$^{e}$ & Empirical & 0.009 & 0.045 & $1-10$ \\
Eureqa (Symbolic Regression)$^{f}$ & Empirical & 0.009 & 0.045 & $10-100$ \\
\hline
{\bf This work$^{\dagger}$} & Empirical & 0.030 & 0.034 & 2.25 (5 PCs) \\
                            & &      &       & 2.90 (4 PCs) \\
Trees for Photo-$z$$^{g,\dagger}$ & Empirical & 0.019 & 0.014 & 0.78 \\
Self-Organizing Maps Photo-$z$ $^{g,\dagger}$ & Empirical & 0.020 & 0.015 & 0.70 \\
Bayesian Photo-{\it z}'s$^{g,\dagger}$ & Template & 0.023 & 0.016 & 1.34 \\
Weighted Average$^{g,\dagger}$ & Hybrid & 0.020 & 0.014 & 0.82 \\
Weighted Average, oracle weighting scheme$^{g,\dagger}$ & Hybrid & 0.019 & 0.014 & 0.67 \\
Weighted Average, shape weighting scheme$^{g,\dagger}$ & Hybrid & 0.019 & 0.014 & 0.81 \\
Weighted Average, fit weighting scheme$^{g,\dagger}$ & Hybrid & 0.020 & 0.014 & 0.90 \\
Bayesian Model Averaging$^{g,\dagger}$ & Hybrid & 0.018 & 0.013 & 0.60 \\
Bayesian Model Combination$^{g,\dagger}$ & Hybrid & 0.018 & 0.013 & 0.59 \\
Hierarchical Bayes$^{g,\dagger}$ & Hybrid & 0.020 & 0.014 & 0.84 \\
\hline
\label{tab:diag}

\end{tabular}
\noindent {\bf Notes.} All codes use the $O^{(a)}$ outlier fraction, unless a dagger ($\dagger$) is placed next to them, for which the $O^{(b)}$ definition is used. $^{a}$ Methods classification: \textit{Empirical}: uses a machine learning technique that usually require supervised or unsupervised learning from a training data set, \textit{Template}: uses synthetic galaxy spectra created from stellar population models, and \textit{Hybrid}: uses  a combination of Template and Empirical methods, applying different weighting schemes to aggregate their results. $^{b}$ Average of $\Delta z$ without the outliers. $^{c}$ Root mean square of $\Delta z$ without the outliers. $^{d}$ Diagnostics presented in \citet{hildebrandt2010}. $^{e}$ \citet{Krone-Martins14a} for $z<0.7$. $^{f}$ \citet{Krone-Martins14a} for $z \geq 0.7$. $^{g}$ Fit diagnostics obtained from \citet{Carrasco14a}.
\end{table*}

\subsection{Real Data}
\label{subsec:realdata}

The SDSS data set was randomly split into training and test sets. The size of the training set was selected such that the catastrophic error plateaued in value.  Initial tests show that the catastrophic errors become stable for training sets that contain more than 500 galaxies and plateaus at around $4,000$ galaxies. We adopt $10,000$ galaxies for the training data set. The PCA decomposition requires only 4 PCAs to retain $99.5\%$ of the variance. This combination of training sample and number of principal components results in a catastrophic error of $O^{(a)}=8.087\%$ within $10$ seconds. Including all 5 components and increasing to a training set size of $50,000$ results in an improvement to $O^{(a)}=6.439\%$. We outline a selection of other diagnostics in Table \ref{tab:diag}.

\subsection{Caveats \& Improvements}
\label{subsec:caveats}

\begin{figure}
   \includegraphics[width=9cm]{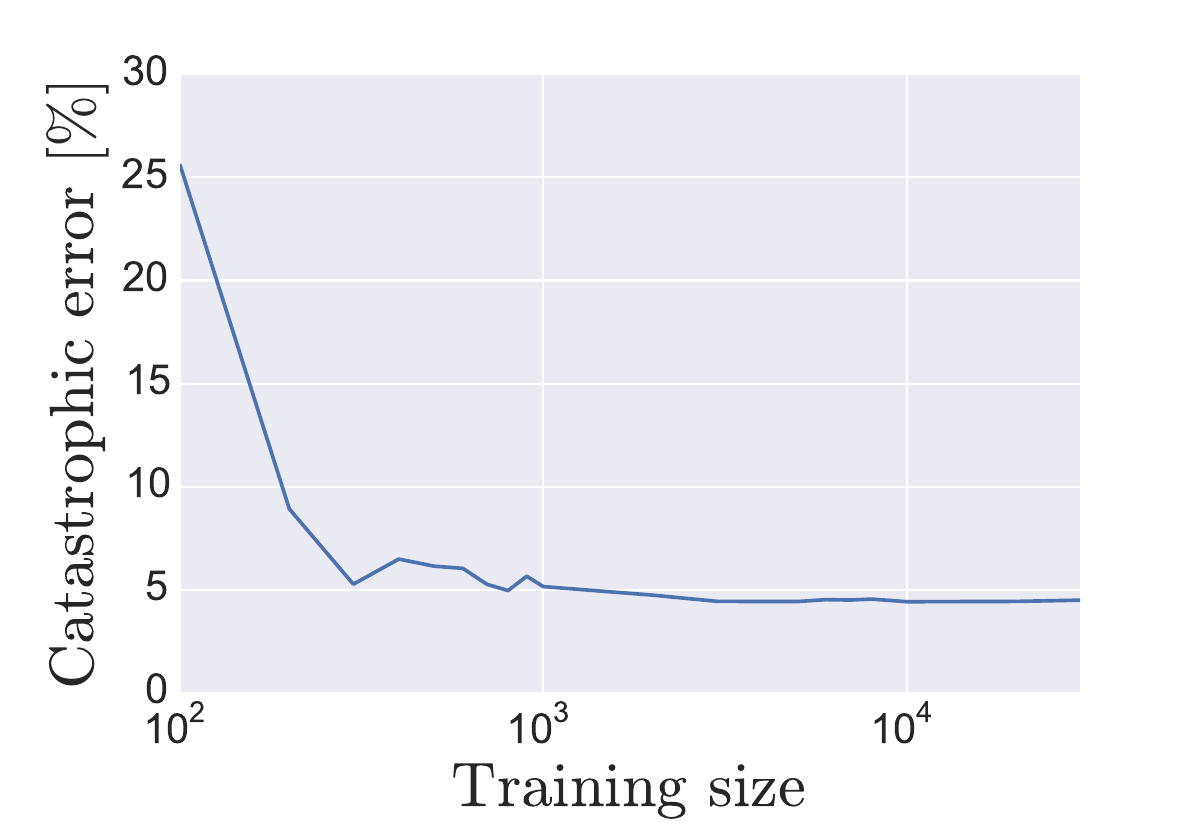}
   \includegraphics[width=9cm]{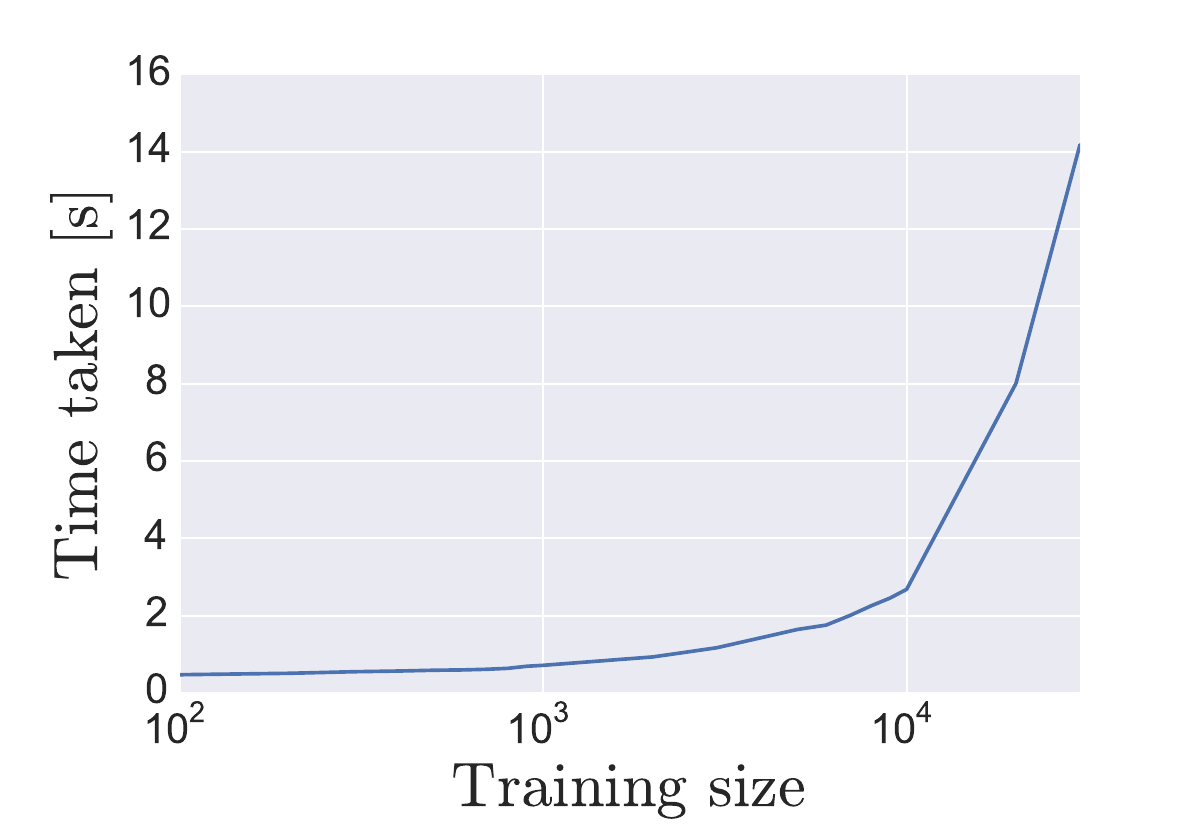}
   \caption{{\bf Top}: Catastrophic error of \photz{} vs. the size of the training set, for the PHAT0 data. {\bf Bottom}: Time taken to complete the GLM fitting as a function of number of elements in the  training set.}
  \label{fig:caveat_trainsize}
\end{figure}

\begin{figure}
   \includegraphics[width=9cm]{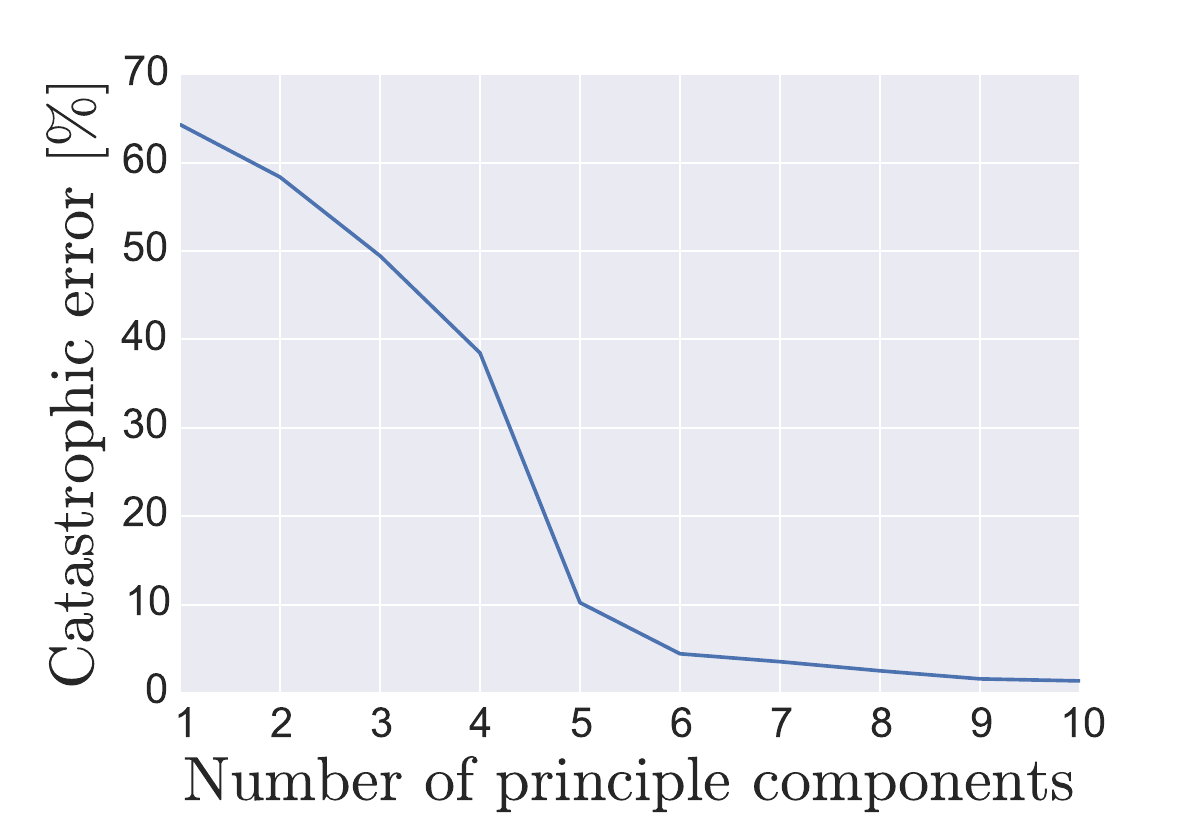}
   \includegraphics[width=9cm]{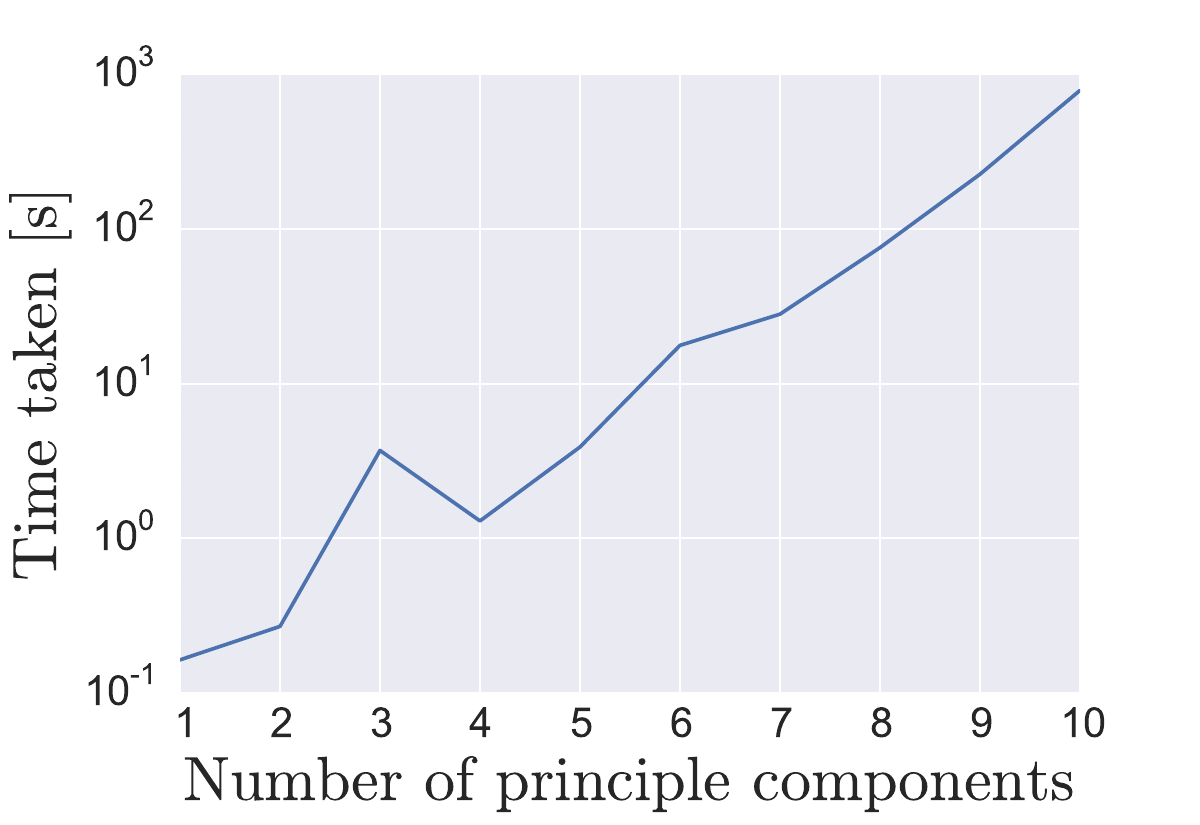}
   \caption{{\bf Top}: Catastrophic error of the fitted \photz{} vs. the number of principal components for the PHAT0 data, considering a fixed training set with 8000 galaxies. {\bf Bottom}: Time taken to complete the GLM fitting as a function of the number of PCs used.}
  \label{fig:caveat_pcasize}
\end{figure}

The use of a GLM to predict \photzs{} can achieve competitive estimates with catastrophic errors of the order of $1-5\%$, comparable with current techniques. It is important to note that there is a wide range of choices that the user can make for their own data set to meet the needs of its use. The most dominant parameters are the size of the training set and the number of principal components used. As one would expect, the time taken increases when both of these parameters are increased and there will be some variation in the resulting catastrophic errors. Figures \ref{fig:caveat_trainsize} and \ref{fig:caveat_pcasize} depict the catastrophic error rate as a function of training set size and number of PCAs used, respectively.

\begin{figure}
   \includegraphics[width=9cm]{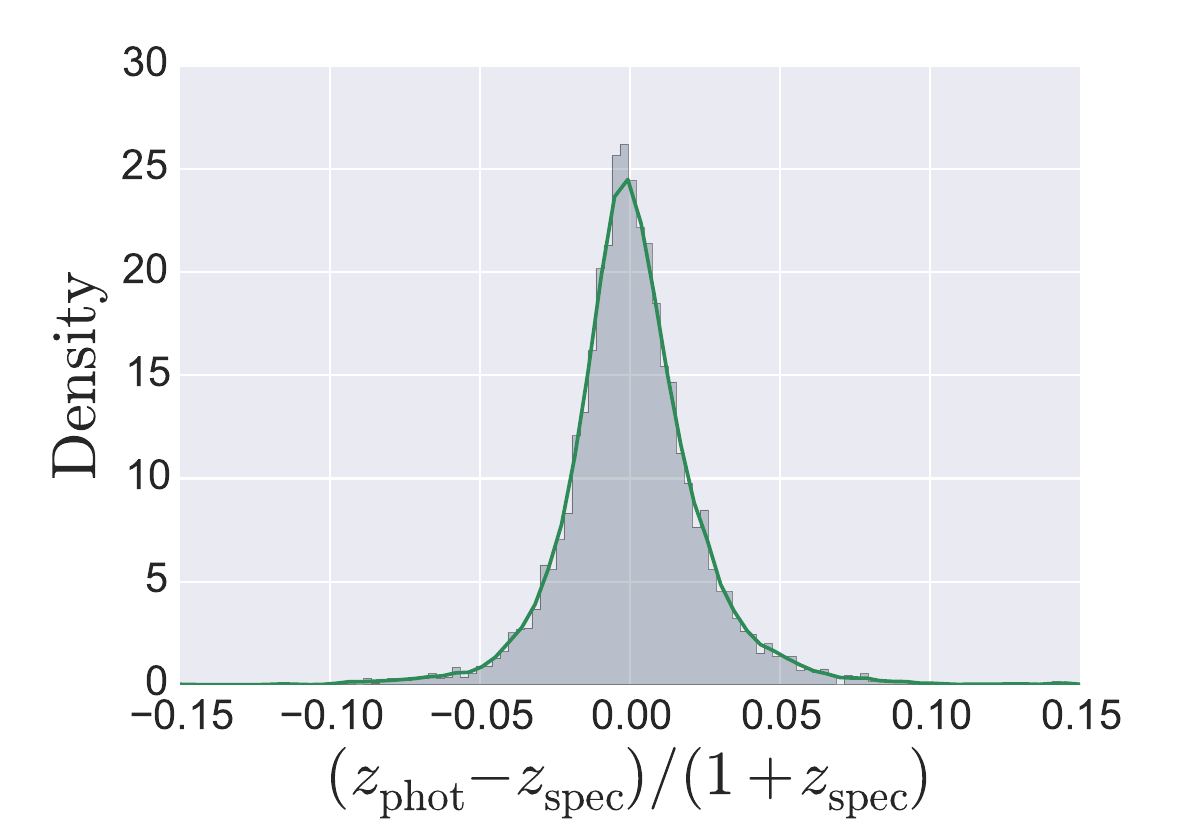}
   \includegraphics[width=9cm]{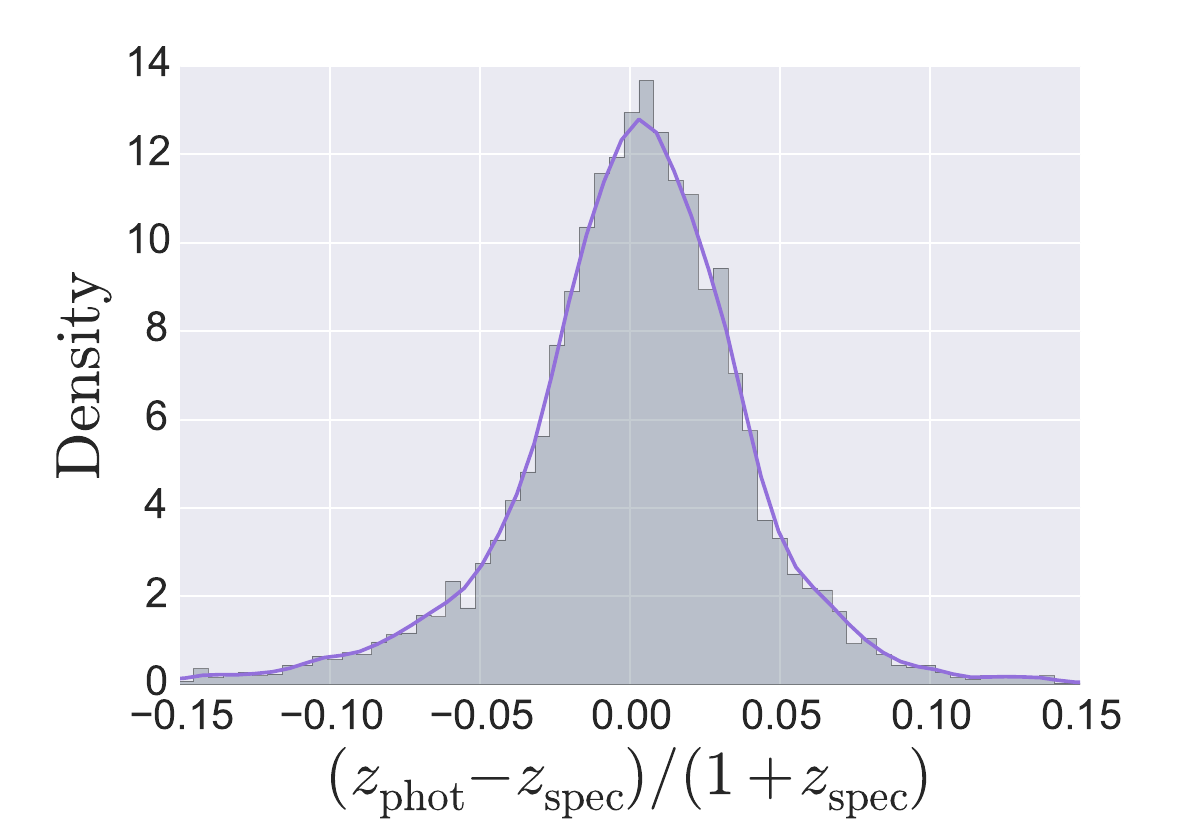}
   \caption{The probability density of the number of outliers. {\bf Top}: Results from  PHAT0 sample (green). {\bf Bottom}: Results from the SDSS sample (purple).}
  \label{fig:kde_plot_1d}
\end{figure}

In addition, as with all ML techniques, one must choose the type of formula used so as to ensure the training data is not over fitted. Increasing the order of the polynomial in equation~\ref{eqn:formula} or including more cross terms will decrease the overall catastrophic error, but will also increase the running time. Such caveats are best left for the user, the catalogue in question, and the needs and requirements of the output photometry. 

Another viable use of GLMs is to make extrapolated predictions of \photz{}s of high-redshift galaxies by learning on low-redshift galaxies. Like other machine learning techniques, GLMs can over-fit the data and make incorrect predictions.
However, GLMs have the advantage that they are not (and should not) be treated as black boxes, like many other empirical and template techniques, e.g. neural networks~\citep{Werner07a}. Instead, the underlying assumptions can be modified, for example: the formula adopted (equation \ref{eqn:formula}), and the underlying effect of the covariates (PCs). As already noted, the relevance of each covariate can be investigated to determine their importance in the model and reveal a physical interpretation. Even though in our analysis we have used PCA to uncorrelate the magnitudes, the resulting influence of each magnitude in the model (the physical interpretation) can still be obtained indirectly from the PCs using techniques such as \textit{projection to latent structures}~\citep{Sasdelli14a}. In combination with a cross-validation process, this would allow the user to optimise the GLM for their needs (e.g. extrapolating predictions) and would highlight the relevance of the covariates in the determination of their wanted values. Furthermore, the extrapolated prediction will still have two inherent gamma features: (i) non-negative values and (ii) heteroscedasticity, consistent with what we expect from the galaxy photometric redshift distributions.

In this study, we have shown that gamma GLMs can speedily and efficiently compute photometric redshifts of galaxies. However, this initial case study could be built upon and improved. This is not within the scope of this paper, but we outline some techniques that could be used to make our methodology better. We emphasis that, although there are a variety of different techniques available to approach the \photz{} problem, GLMs provide a straightforward and statistically coherent way of including the modelling (prior) within the regression process. This is done through the choice of the family distribution. Traditional regression methods can sometimes yield negative, non-physical values, which we have avoided from the start by using the gamma family distributions. Thus, the enhancements cited bellow are all within this framework.

\begin{figure}
  \includegraphics[width=9cm]{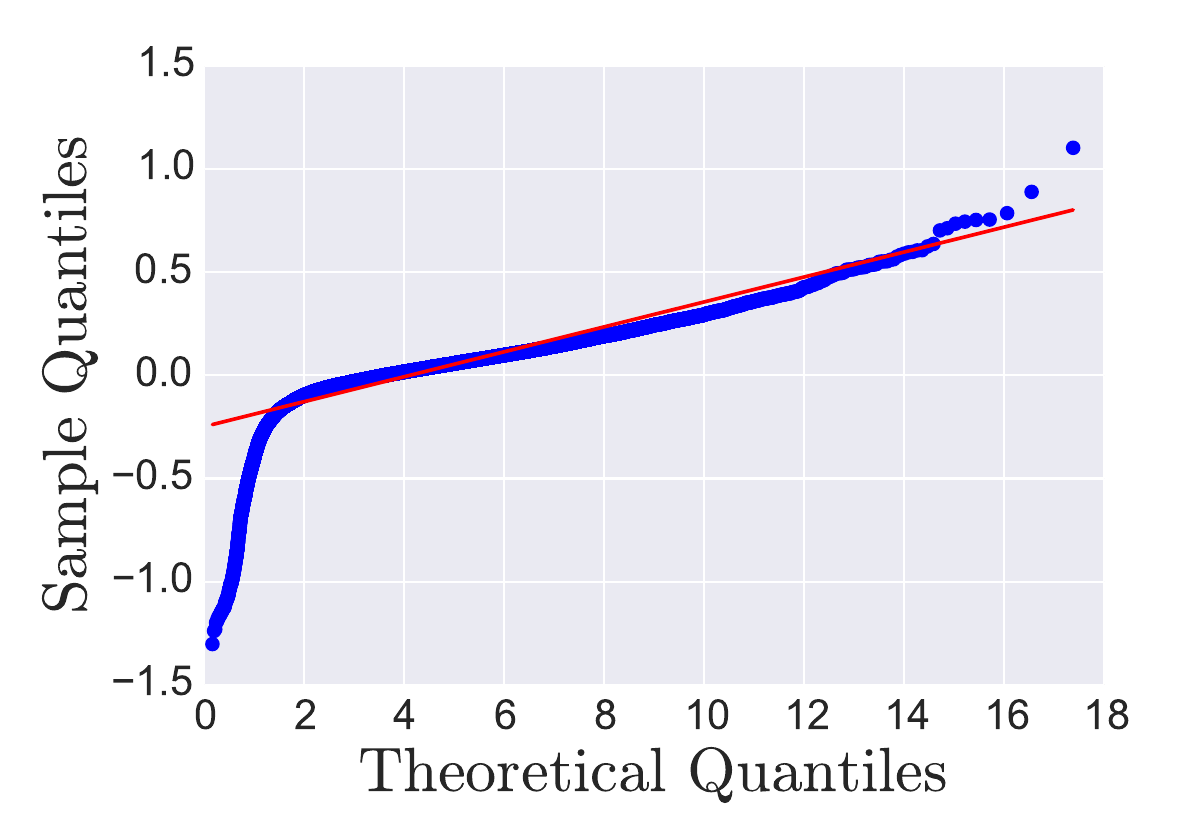}
  \caption{A Q-Q plot of the residuals of the best fit gamma model compared to a gamma distribution. For visual reasons we have clipped one outlier at $y=6.5$. See the text for an interpretation of the plot.}
  \label{fig:qqplot}
\end{figure}

To determine how to further develop this methodology, we utilise a standard diagnostic, in statistics, called the Q-Q (quantile-quantile) plot. We compare the residuals of the best-fit gamma model of the PHAT0 data set to a gamma distribution, as seen in Fig. \ref{fig:qqplot}. When the two distributions are similar it is expected that a straight line will go through the data set, as is seen in our case (red line). As the best-fit line is shallower than a normal $y=x$ line, this implies that the theoretical gamma distribution is more dispersed than the sample PHAT0 distribution. There is a deviation at the lower quartiles ($<1$) that suggests the residuals are deviating from a gamma. This skewing towards lower quantiles is alleviated when selecting galaxies that lie in the redshift range $0.25<z<1.0$. One way to address this problem would be to consider more complex models such as generalized additive models for location, scale and shape~\citep[GAMLSS;][]{Stasinopoulos07a}. GAMLSS are more lenient and do not  require an exponentiated family distribution to be used, but instead a general distribution family that can be highly skewed, kurtotic, and/or continuous and discrete. Investigation of such models is not within the scope of this paper, but are expected to be addressed  in future versions of the code.

\section{Conclusions}
\label{sec:conclusion}

Generalized linear models are widely used through a multitude of academic disciplines, but have been relatively untouched throughout the astronomical community. Their straight forward implementation and possibility of allowing physically relevant  predictions make GLMs a great candidate for competing with conventional methods of modelling that are more often turning to techniques involving ML, and more specifically neural networks. For the case of photometric redshifts, an adoption of a gamma family reflects two important characteristics of the data: (i) a non-negative and continuous measurement, and (ii) heteroscedasticity, i.e., the variance of the photo-$z$ measurements changes according to the redshift. The gamma GLM intrinsically assumes that higher values of photo-$z$ have a larger intrinsic scatter.

Upcoming wide field sky surveys, such as the LSST, will take the challenge of determining \photz{}s to an unprecedented scale. To this end, we have outlined the use of GLMs to tackle the problem of estimating \photz{}s for large samples of galaxies from their multi-band photometry in a semi-supervised learning manner. We demonstrate that GLMs can be trained on $50,000$ galaxies with $5$ principle components in $3$ seconds and $10$ principle components in $10$ minutes, using a standard laptop (AMD Athlon X2 Dual-Core QL-64 processor with $1.7\, \rm GB$ RAM on the Ubuntu 10.04 OS),  and can reach catastrophic errors of $1-5\%$, comparable with current techniques involving template fitting or ML.

In summary, GLMs offer a simple and efficient way of tackling many problems within astronomy that are usually computationally heavy or require large training samples. To promote their use within the astronomical community we have developed a suite of libraries and a web application to allow GLMs to be used to determine \photzs{} from a user's own galaxy sample with a simple click of a button. 



\section*{Acknowledgments}
We thank V. Busti, E. D. Feigelson, M. Killedar, J. Buchner,  and A. Trindade for interesting suggestions and comments. 
JE, RSS and EEOI thank the SIM Laboratory of the \emph{Universidade de Lisboa} for hospitality during the development of this work. 
Cosmostatistics Initiative (COIN)\footnote{\url{https://asaip.psu.edu/organizations/iaa/iaa-working-group-of-cosmostatistics}} is a non-profit organisation whose aim is to nourish the synergy between astrophysics, cosmology, statistics and machine learning communities.
This work was partially supported by the ESA VA4D project (AO 1-6740/11/F/MOS). AKM thanks the Portuguese agency \emph{Funda\c c\~ao para Ci\^encia e Tecnologia} -- \emph{FCT}, for financial support (SFRH/BPD/74697/2010). EEOI is partially supported by the Brazilian agency CAPES (grant number 9229-13-2). Work on this paper has substantially benefited from using the collaborative website AWOB\footnote{\url{http://awob.mpg.de}} developed and maintained by the Max-Planck Institute for Astrophysics and the Max-Planck Digital Library. This work was written on the collaborative \texttt{WriteLatex} platform\footnote{\url{www.writelatex.com}}, and made use of the GitHub\footnote{\url{www.github.com}} a web-based hosting service and \texttt{git} version control software. This work made use of the cloud based hosting platform \texttt{ShinyApps.io}\footnote{\url{http://www.shinyapps.io}}. This work used the following public scientific Python packages \texttt{scikit-learn} \texttt{v0.15}\footnote{\url{github.com/scikit-learn}}~\citep{Pedregosa11a}, \texttt{seaborn v0.3.1}\footnote{\url{github.com/mwaskom/seaborn}}, and \texttt{statsmodels v0.6.0}\footnote{\url{github.com/statsmodels}}. Funding for SDSS-III\footnote{\url{http://www.sdss.org}} has been provided by the Alfred P. Sloan Foundation, the Participating Institutions, the National Science Foundation, and the U.S. Department of Energy Office of Science.



\bibliography{glm_photoz}
\bibliographystyle{model2-names}



\appendix
\label{sec:appendix}

\section{R package}
\label{app:R}

The R package is publicly available.
It can be obtained either through CRAN (Comprehensive R Archive Network) using the package name \texttt{CosmoPhotoz}, or through the COIN GitHub repository\footnote{\url{https://github.com/COINtoolbox/COSMOPhotoz/}}. The stable CRAN release can be easily installed from within R using the standard function \texttt{install.packages()}, while the GitHub version can be installed via the function \texttt{install\_github()} from the package \texttt{devtools}.

There are two ways to perform the photometric redshift estimation using the R package. The simplest, but less flexible, way is to perform a direct call to the \texttt{CosmoPhotoZestimator()} function using two data frames: one containing the data adopted for training, and another containing the photometric data for photometric redshift estimation. The code bellow shows how to perform the redshift estimate using the PHAT0 data included in the package.

\lstset{
	language=R,
	keywordstyle=\bfseries\ttfamily\color[rgb]{0,0,1},
	identifierstyle=\ttfamily,
	commentstyle=\color[rgb]{0.133,0.545,0.133},
	stringstyle=\ttfamily\color[rgb]{0.627,0.126,0.941},
	showstringspaces=false,
	basicstyle=\footnotesize,
	numberstyle=\tiny,
	numbers=left,
	stepnumber=1,
	numbersep=8pt,
	tabsize=2,
	breaklines=true,
	prebreak = \raisebox{0ex}[0ex][0ex]{\ensuremath{\hookleftarrow}},
	breakatwhitespace=false,
	aboveskip={1.5\baselineskip},
  columns=fixed,
  upquote=true,
  extendedchars=true,
 backgroundcolor=\color{gray!5},
}
\begin{lstlisting}
  library(CosmoPhotoz)
  data(PHAT0train)#Data for training
  data(PHAT0test)#Data for estimation
  # Run the analysis 
  photoZest <- CosmoPhotoZestimator(PHAT0train,
  PHAT0test, numberOfPcs=6)
  # Not using robust PCA is considerably faster, 
  # but the results are worse
  photoZestN <- CosmoPhotoZestimator(PHAT0train, 
  PHAT0test,  numberOfPcs=6, 
  robust=FALSE) 
  # Create a boxplot showing the results
  plotDiagPhotoZ(photoz = photoZest, 
  specz = PHAT0test$redshift, 
  type = "box")
\end{lstlisting} 

The most flexible way to use the package, however, is to perform a step by step analysis using the individual functions provided. The following code exemplifies how it is possible to perform the redshift estimate using some of such functions.

\begin{lstlisting}
  library(CosmoPhotoz)
  data(PHAT0train)# Data for training
  data(PHAT0test)# Data for estimation
  
  # Combine the training and test data and 
  # calculate the principal components
  PC_comb <- computeCombPCA(
  subset(trainData,select=c(-redshift)),
  subset(testData, select=c(-redshift)),
  robust=TRUE)
  Trainpc <- cbind(PC_comb$x, 
  redshift=trainData$redshift)
  Testpc <- PC_comb$y
  
  # Formula based on the PCs
  formM <- redshift~poly(Comp.1,2)*
  poly(Comp.2,2)*Comp.3*Comp.4*
  Comp.5*Comp.6
          
  # GLM fitting
  Fit <- glmTrainPhotoZ(Trainpc, formula=formM, 
  method="Bayesian", family="gamma")
          
  # Photo-z estimation
  photoZtest <- glmPredictPhotoZ(Fit$glmfit, newdata=Testpc, 
  type="response")
  
  # Print Photo-z estimation
   print(photoZtest$photoz)
  
  # Estimate confidence intervals errors  
  photoz_temp <- predict(Fit$glmfit, newdata=Testpc, type="link", se.fit = TRUE)
  photoz <- photoz_temp$fit
  
  critval <- 1.96 ## approx 95% Confidence Interval
  upr <- photoz_temp$fit + (critval * photoz_temp$se.fit)
  lwr <- photoz_temp$fit - (critval * photoz_temp$se.fit)
  fit <-  photoz_temp$fit
  
  fit2 <- Fit$glmfit$family$linkinv(fit)
  upr2 <- Fit$glmfit$family$linkinv(upr)#upper limit
  lwr2 <- Fit$glmfit$family$linkinv(lwr)#lower limit
  
  
  print(upr2)# upper limit
  print(lwr2)# lower limit
    
  # Create a boxplot showing the results
  plotDiagPhotoZ(photoz = photoZtest$photoz, 
  specz = PHAT0test$redshift, 
  type = "box")
\end{lstlisting}           

The functions provided by the package and that are visible to the user are: 

\begin{enumerate}[1.]
  \item \texttt{computeCombPCA()}
  Computes PCA projections of the combined data sets.
  
  \item \texttt{computeDiagPhotoZ()}
   Computes a list of summary statistics of the redshift estimation (mean, sd, median, mad, outliers).

  \item \texttt{CosmoPhotoZestimator()}
   Computes redshift estimates from photometric data and a training data set with photometry and spectroscopy. The estimation is based on GLMs.
  
  \item \texttt{glmPredictPhotoZ()}
Predicts photometric redshifts   for  a given a GLM fit object.
  
  \item \texttt{glmTrainPhotoZ()}
  Fits a GLM for photometric redshift estimation. A Bayesian fit or a normal fit may be adopted, using the link functions gamma and inverse Gaussian. 
  
  \item \texttt{plotDiagPhotoZ()}
  Plots diagnostics for redshift estimations. The following types of plot are available: 1D kernel density estimation of the errors (\texttt{errordist}), observed versus predicted 2D density plot (\texttt{predobs}), a violin and a box plot with errors at redshift bins (\texttt{errorviolins} or \texttt{box}).
   
\end{enumerate}

The packages additionally includes two dataframes, \texttt{PHAT0train} and \texttt{PHAT0test}, containing 161042 and 8478 objects, and comprising 12 variables (11 bands plus the redshift).

A detailed reference manual of the package can be found at the documentation in CRAN webpage or inside the package. 

\subsection{Shiny Package}
\label{app:shiny}

The R package is also accompanied by a shiny\footnote{\url{http://shiny.rstudio.com/}} application that can be hosted locally or deployed by the user at a webserver. This application allows the user to run the photometric redshift estimation, to configure many parameters of the code visually and experiment with the results. It also allows the user to either use the PHAT0 data, or to upload data files (the expected format can be found at the application’s help tab). A screenshot of this application can be seen in Fig. \ref{fig:shinyAppScreenshot}.

To run the interface, after installing the package it is necessary to use the following command: 
  
 \begin{lstlisting} 
 runApp(paste(find.package("CosmoPhotoz"),"/CosmoPhotoz_shiny",sep=""))
\end{lstlisting}

\begin{figure*}
\centering
   \includegraphics[width=0.925\textwidth]{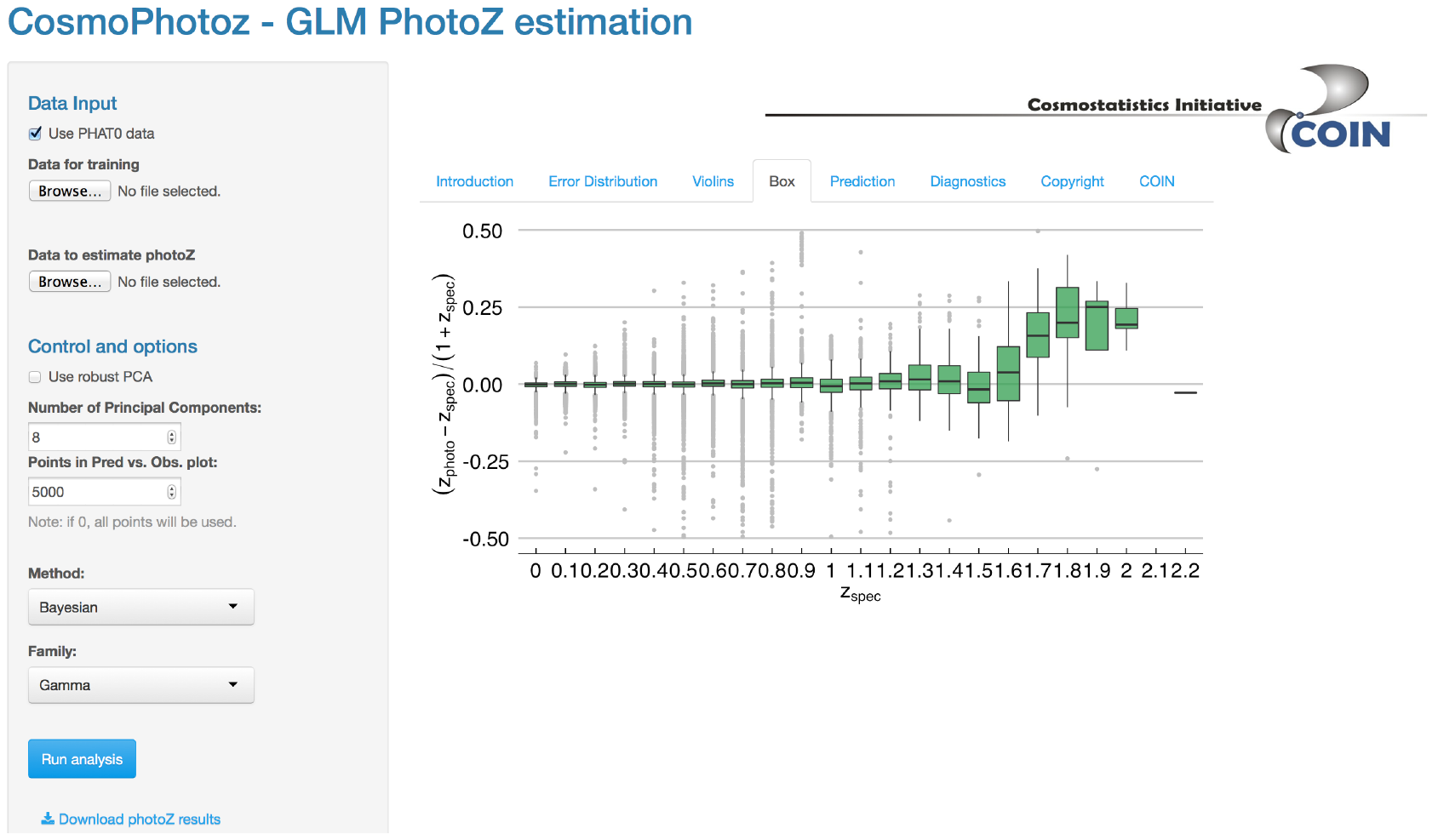}
   \caption{A screenshot of the Shiny web application running on the \texttt{shinyapps.io} cloud. In this screenshot the application was estimating redshifts for a subset of the PHAT0 data set. This application is publicly available at \url{https://cosmostatisticsinitiative.shinyapps.io/CosmoPhotoz}.}
  \label{fig:shinyAppScreenshot}
\end{figure*}

Finally, this application can also be used via the web. This option requires no local installation, but the actual processing may be slower. This web interface is hosted by the shinyapps.io platform\footnote{\url{http://www.shinyapps.io}}, and can be accessed directly at \url{https://cosmostatisticsinitiative.shinyapps.io/CosmoPhotoz}.

\section{Python Package}
\label{app:python}

The Python package is publicly available. The source code can be accessed directly at the COIN GitHub repository. The package can also be easily installed via the Python Package Index\footnote{\url{https://pypi.python.org/pypi/CosmoPhotoz/0.1}} using \texttt{easy\_install} or \texttt{pip}. Complete documentation is fully accessible from the \textit{Read-the-docs}\footnote{\url{http://cosmophotoz.readthedocs.org}} platform. The package can be used in two ways. The first is a binary file that can be run on the command line with little interaction required:

\lstset{
	language=Bash,
	keywordstyle=\bfseries\ttfamily\color[rgb]{0,0,1},
	identifierstyle=\ttfamily,
	commentstyle=\color[rgb]{0.133,0.545,0.133},
	stringstyle=\ttfamily\color[rgb]{0.627,0.126,0.941},
	showstringspaces=false,
    basicstyle=\footnotesize,
	numberstyle=\tiny,
	numbers=left,
	stepnumber=1,
	numbersep=8pt,
	tabsize=2,
	breaklines=true,
	prebreak = \raisebox{0ex}[0ex][0ex]{\ensuremath{\hookleftarrow}},
	breakatwhitespace=false,
	aboveskip={1.5\baselineskip},
  columns=fixed,
  upquote=true,
  extendedchars=true,
 backgroundcolor=\color{gray!5},
}

  \begin{lstlisting}[language=bash] 
    run_glm.py --dataset sample.csv
    
    run_glm.py --dataset train.csv test.csv
    
    run_glm.py --dataset sample.csv
               --num_components 3
               --training_size 10000
               --family Gamma
               --link log
\end{lstlisting}

The library can also be imported in the standard way by the user. There is a \texttt{PhotoSample} class that can be fully manipulated. We give a few examples of how the class can be utilised for personal use with photometric catalogues. Firstly, just executing the analysis like the binary.

\lstset{
	language=Python,
	keywordstyle=\bfseries\ttfamily\color[rgb]{0,0,1},
	identifierstyle=\ttfamily,
	commentstyle=\color[rgb]{0.133,0.545,0.133},
	stringstyle=\ttfamily\color[rgb]{0.627,0.126,0.941},
	showstringspaces=false,
	basicstyle=\footnotesize,
	numberstyle=\tiny,
	numbers=left,
	stepnumber=1,
	numbersep=8pt,
	tabsize=2,
	breaklines=true,
	prebreak = \raisebox{0ex}[0ex][0ex]{\ensuremath{\hookleftarrow}},
	breakatwhitespace=false,
	aboveskip={1.5\baselineskip},
  columns=fixed,
  upquote=true,
  extendedchars=true,
 backgroundcolor=\color{gray!5},
}

\begin{lstlisting}
 from CosmoPhotoz.photoz import PhotoSample
 # Use PHAT0 catalogue supplied
 # with the software
 UserCatalogue = PhotoSample(filename="PHAT0",\
 family="Gamma", \link="log")
 UserCatalogue.run_full()
\end{lstlisting}

Secondly, selecting a log link and only making the violin plot.

\begin{lstlisting}
 # Import the PhotoSample class
 from CosmoPhotoz.photoz import PhotoSample
 # Instantiate an object of PhotoSample class
  UserCatalogue = PhotoSample(filename="PHAT0",\
  family="Gamma", \
  link="log")

 # Select the link
  UserCatalogue.link = "log"
    
 # Carry out PCA
  UserCatalogue.do_PCA()
    
 # Split into training and test data
  UserCatalogue.split_sample(random=True)    
    
 # GLM fitting and photo-z estimation
  UserCatalogue.do_GLM()
    
 # Create a violin plot showing the results
  UserCatalogue.make_violin()
\end{lstlisting}

Finally, to determine the number galaxies required to achieve at least a catastrophic error of $5.937\%$.

\begin{lstlisting}
 # Import the PhotoSample class
 from CosmoPhotoz.photoz import PhotoSample
 import numpy as np

 #Instantiate the class
 UserCatalogue = PhotoSample(filename="PHAT0",\
 family="Gamma", \link="log")

 # Make a training size array to loop through
 train_size_arr = np.arange(500,10000,500)
 catastrophic_error = []

 # Select your number of components
 UserCatalogue.num_components = 4

 for i in range(len(train_size_arr)):
 UserCatalogue.do_PCA()
 UserCatalogue.test_size = train_size_arr[i]
 UserCatalogue.split_sample(random=True)
 UserCatalogue.do_GLM()
 catastrophic_error.append(\
 UserCatalogue.catastrophic_error)

 min_indx = np.array(catastrophic_error) < 5.937
 optimum_train_size = train_size_arr[min_indx]
    
 # Print the output to the user
 print(optimum_train_size)
\end{lstlisting}

The main methods of the \texttt{PhotoSample} class are the following:
\begin{enumerate}[1.]
  \item \texttt{\_\_init\_\_()}
  Constructor of the class. This is used to define the public attributes of the class, e.g., number of PCA components, size of the training sample, file name, etc.
  
  \item \texttt{do\_PCA()}
  Carries out PCA on the data set. Using the variance per principal component, it determines the optimal number of principal components ensuring that 99.95\% of the variance is retained.
  
  \item \texttt{split\_sample()}
  Randomly splits the sample if the user gives a single file, for which they wish to use a random sample to train and test the GLM. The method also reconstructs a final {\bf\texttt{Pandas DataFrame}} object to be used in the GLM fitting.
  
  \item \texttt{do\_GLM()}
  Instantiates the GLM class and selects the family and link type to be used. If the user does not interfere, the gamma family and inverse power link will be used, which may not be fully optimal in all situations. 
  
  \item \texttt{make\_kde\_1d()}
  Creates a plot that shows the probability density function of the outlier fraction.
  
  \item \texttt{make\_kde\_2d()}
  Creates a plot that shows the probability density function of the photometric and spectroscopic redshifts on the upper and right most axis. In the centre a 2D probability density plot is shown.

  \item \texttt{make\_violin()}
  Creates a violin plot. A violin plot shows the probability density of the outlier fraction for a given redshift interval.
  
  \item \texttt{write\_to\_file()}
  Writes the GLM \photz{} predictions to a file named \emph{glmPhotoZresults.csv} when the user supplies a \emph{filename\_test} and \emph{filename\_train}{\bf.}
  
\end{enumerate}

We note that there are slightly differences to how the \texttt{R} and \texttt{Python} codes solve the SVD matrices in the PCA step and also how they minimise the GLM. We find that the predicted redshifts from both packages have a Pearson-R value of 1, and have a spread that is of the order and smaller than the intrinsic spread of the predicted vs. measured redshifts, and are thus negligible.

\section{SDSS SQL Photometry Query}
\label{app:sql}
The following SQL code can be run on the SDSS CasJobs service \footnote{\url{http://skyserver.sdss3.org/CasJobs/}} to obtain the photometry of the galaxy sample investigated within this article.

\begin{lstlisting}[language=sql]

  SELECT s.specObjID, g.dered_u, g.dered_g, g.dered_r, g.dered_i, g.dered_z, s.z AS redshift INTO mydb.specObjAllz_cleanphoto
  FROM SpecObj As s JOIN Galaxy as g ON s.specobjid = g.specobjid, PhotoObj
  WHERE class = "GALAXY" 
      AND zWarning = 0
      AND g.objID = PhotoObj.ObjID
      AND PhotoObj.CLEAN=1
      AND s.z > 0
      AND g.dered_u > 0
      AND g.dered_g > 0
      AND g.dered_r > 0
      AND g.dered_i > 0
      AND g.dered_z > 0
\end{lstlisting}


\end{document}